
\input phyzzx

\def\math{\mathsurround 0pt}
\def\oversim#1#2{\lower.5pt\vbox{\baselineskip0pt \lineskip-.5pt
        \ialign{$\math#1\hfil##\hfil$\crcr#2\crcr{\scriptstyle\sim}\crcr}}}

\def\({\left(} \def\){\right)}
\def\[{\left[} \def\]{\right]}
\def\pa{\partial}
\def\frac#1#2{
  {\mathchoice{{\textstyle{#1\over #2}}}{#1\over #2}{#1\over #2}{#1 \over #2}}}
\def\half{{\mathchoice{{\textstyle{1\over 2}}}{1\over 2}{1\over 2}{1 \over 2}}}

\def\overleftrightarrow#1{\vbox{\ialign{##\crcr
    $\scriptstyle\leftrightarrow$\crcr\noalign{\kern0.3pt\nointerlineskip}
    $\hfil\displaystyle{#1}\hfil$\crcr}}}
\def\dbw{\overleftrightarrow\partial}

\def\al{\alpha}

\def\ga{\gamma}

\def\th{\theta}

\def\la{\lambda}

\def\si{\sigma}

\def\De{\Delta}

\def\La{\Lambda}

\def\na{\nabla}

\def\vp{\varphi}

\def\tn{t_{\rm n}}
\def\Gl{G_{\rm l}}
\def\ms{m_{\rm s}}
\def\mv{m_{\rm v}}

\REF\one{T.W.B. Kibble, {\it J. Phys.} {\bf A9}, 1387 (1976).}
\REF\two{A. Vilenkin, {\it Phys. Rep.} {\bf 121}, 263 (1985).}
\REF\three{N. Turok, in {\it Particles, Strings and Supernovae} eds.
A. Jevicki and C.-I.~Tan (World Scientific, Singapore, 1989).}
\REF\four{N. Mermin, {\it Rev. Mod. Phys.} {\bf 51}, 591 (1979).}
\REF\five{M. Salomaa and G. Volovik, {\it Rev. Mod. Phys.} {\bf 59},
533 (1987).}
\REF\six{P. de Gennes, ``The Physics of Liquid Crystals" (Clarendon
Press, Oxford, 1974).}
\REF\seven{A. Abrikosov, {\it JETP} {\bf 5}, 1174 (1957).}
\REF\eight{Ya.B. Zel'dovich and M. Khlopav, {\it Phys. Lett.} {\bf
79B}, 239 (1978); \nextline
J. Preskill, {\it Phys. Rev. Lett.} {\bf 43}, 1365 (1979).}
\REF\nine{R. Brandenberger, A.-C. Davis and M. Hindmarsh, {\it Phys.
Lett.} {\bf 263}, 239 (1991); \nextline
R. Brandenberger and A.-C. Davis, {\it Phys. Lett. B}, in press (1993).}
\REF\ten{Ya.B. Zel'dovich, {\it Mon. Not. R. astron. Soc.} {\bf 192},
663 (1980); \nextline
A. Vilenkin, {\it Phys. Rev. Lett.} {\bf 46}, 1169 (1981); \nextline
R. Brandenberger, {\it Phys. Scripta} {\bf T36}, 114 (1991).}
\REF\eleven{D. Bennett and S. Rhie, {\it Phys. Rev. Lett.} {\bf 65},
1709 (1990).}
\REF\twelve{N. Turok, {\it Phys. Rev. Lett.} {\bf 63}, 2625
(1989);\nextline
N. Turok, {\it Phys. Scripta} {\bf T36}, 135 (1991).}
\REF\thirteen{I. Chuang, R. Durrer, N. Turok and B. Yurke, {\it
Science} {\bf 251}, 1336 (1991); \nextline
M. Bowick, L. Chandar, E. Schiff and A. Srivastava, Syracuse preprint
SU-HEP-4241-512 (1992).}
\REF\fourteen{J. Ye and R. Brandenberger, {\it Mod. Phys. Lett.} {\bf
A5}, 157 (1990); \nextline
A. Srivastava, {\it Phys. Rev.} {\bf D45}, 3304 (1992).}
\REF\fifteen{J. Ye and R. Brandenberger, {\it Nucl. Phys.} {\bf B346},
149 (1990).}
\REF\sixteen{S. Rudaz and A. Srivastava, Univ. of Minnesota preprint
UMN-TH-1028-92 (1992).}
\REF\seventeen{T. Vachaspati and A. Achucarro, {\it Phys. Rev.} {\bf
D44}, 3067 (1991).}
\REF\eightteen{T. Vachaspati, {\it Phys. Rev. Lett.} {\bf 68}, 1977
(1992).}
\REF\MarRus{J. March-Russell, {\it Phys. Lett.} {\bf B296}, 364 (1992).}
\REF\EffPot{D. A. Kirzhnits and A. D. Linde, {\it JETP} {\bf 40}, 628 (1974);
\nextline
S. Weinberg, {\it Phys. Rev.} {\bf D9} 3320 (1974);
\nextline
L. Dolan and R. Jackiw, {\it Phys. Rev.} {\bf D9} 3320 (1974);
\nextline
D. A. Kirzhnits and A. D. Linde, {\it Ann. Phys.} {\bf 101},
 {195} (1976).}
\REF\nineteen{M.B. Voloshin, I.Yu. Kobzarev and L.B. Okun, {\it
Yad. Fiz.} {\bf 20}, 1229 (1974) [{\it Sov. J. Nucl. Phys.}
{\bf 20}, 644 (1975)];\nextline
S. Coleman, {\it Phys. Rev.} {\bf D15}, 2929 (1977);
\nextline
C. Callan and S. Coleman, {\it Phys. Rev.} {\bf D16}, 1762 (1977).}
\REF\Lin{A. D. Linde, {\it Nucl. Phys.} {\bf B216}, 421 (1983).}
\REF\twenty{T. Prokopec, {\it Phys. Lett.} {\bf B262}, 215 (1991);
\nextline
R. Leese and T. Prokopec, {\it Phys. Rev.} {\bf D44}, 3749 (1991).}
\REF\twentyone{S. Hawking, I. Moss and J. Stewart, {\it Phys. Rev.}
{\bf D26}, 2681 (1982).}
\REF\twentythree{H. Nielsen and P. Olesen, {\it Nucl. Phys.} {\bf
B61}, 45 (1973).}
\REF\twentyfour{M. Hindmarsh, {\it Phys. Rev. Lett.} {\bf 68}, 1263
(1992).}
\REF\twentyfive{M. Hindmarsh, {\it Nucl. Phys.} {\bf B392} 461 (1993).}
\REF\AreaProb{{\it Mathematical questions with their solutions} vol.
\romannumeral8\ p. 21, ed. W. J. Miller (Hodgson and Son, London, 1868).}
\REF\twentysix{M. James, L. Perivolaropoulos and T. Vachaspati, {\it
Phys. Rev.} {\bf D46}, R5232 (1992).}
\REF\HolPlus{R. Holman, S. Hsu, T. Vachaspati and R. Watkins,
{\it Phys. Rev.} {\bf D47}, 5352 (1992).}
\REF\twentyseven{M. Dine et al, {\it Phys. Rev.} {\bf D46}, 550 (1992).}

\FRONTPAGE
\line{\hfill DAMTP-93-13}
\line{\hfill BROWN-HET-902}
\line{\hfill {\tt hep-ph/9307203}}
\line{\hfill June 1993}
\vskip1.5truein
\titlestyle{{ FORMATION OF TOPOLOGICAL DEFECTS IN FIRST
ORDER PHASE TRANSITIONS }}
\bigskip
\author{Mark Hindmarsh$^{a,c)}$\footnote{}{
\hskip -7pt\llap{$^{c)}$}\hskip 2pt
Address after October 1st 1993:  School of Mathematical and
Physical Sciences, University of Sussex, Brighton BN1 9QH, U.K.},
Anne-Christine Davis$^{a)}$
and Robert Brandenberger$^{b)}$}
\bigskip
\centerline{$^{a)}$ {\it Department of Applied Mathematics and
Theoretical Physics} }
\centerline{\it University of Cambridge, Cambridge CB3 9EW, U.K.}
\vskip.3cm
\centerline{ $^{b)}$ {\it Department of Physics}}
\centerline{{\it Brown University, Providence, RI 02912, USA}}
\bigskip
\abstract
We analyze the evolution of scalar and gauge fields during first
order phase transitions and show how the Kibble
mechanism$^{\one)}$ for the formation of topological defects emerges from
the underlying dynamics, paying  particular attention
to problems posed by gauge
invariance when a local symmetry is
spontaneously broken.  We discuss also the application of the mechanism to
semilocal defects and electroweak strings.

\endpage
\chapter{Introduction}

Topological defects arising in theories with a spontaneously broken global or
local symmetry  play an  important role in various branches
of physics from cosmology$^{\two,\three)}$ to condensed matter
physics$^{\four-\seven)}$.

In cosmology, topological defects are expected to form during phase
transitions in the early Universe.  It is important to know the number density
of defects at the time of formation, as well as other statistics such as
correlation functions.  Depending on the initial number density,
many physics models admitting local monopoles
or domain walls are ruled out$^{\eight)}$. For defect-induced
baryogenesis$^{\nine)}$
it is important to know the initial separation of defects, and for structure
formation scenarios using topological defects -- cosmic strings$^{\ten)}$,
global monopoles$^{\eleven)}$ or global textures$^{\twelve)}$ -- it is also
useful to
understand the initial distribution of defects.

In 1976, Kibble$^{\one)}$ suggested a simple mechanism for the formation of
defects in cosmological phase transitions, where the order parameter is a
scalar quantum field $\phi$ (which may be elementary or composite).
It has two key ingredients: randomness of the phases of the scalar
field $\phi$ on length scales larger than some correlation length $\xi$, and
the geodesic rule for interpolating the values of $\phi$ on curves connecting
points in different correlation volumes.  The first assumption states that
$\phi$ takes on random values in its vacuum manifold ${\cal M}$ at points
separated by a distance greater than $\xi$, the second one states that along a
curve in space connecting two such points, the field $\phi$ traces out a
geodesic path on the vacuum manifold.

For global defects, the Kibble mechanism has been verified in liquid crystal
experiments$^{\thirteen)}$ and in numerical simulations in field
theory$^{\fourteen)}$.  For
local defects, convincing experiments are lacking (see, however, Refs.
\fifteen\ for
some numerical results for local vortex formation).

Recently, Rudaz and Srivastava$^{\sixteen)}$ have argued that in local field
theories, the geodesic rule is not justified and that the rate of defect
formation might be much smaller than what would be obtained by naively
applying the Kibble mechanism.  The core of the objection is that the Kibble
mechanism is formulated in a gauge--dependent way, and that a geodesic curve
in the vacuum manifold of $\phi$ does not necessarily minimize the energy
density and hence should not play a distinguished role.
These objections  highlight the need to reconsider
the Kibble mechanism for defect formation in theories with local symmetry
breaking.

In this paper we reconsider the Kibble mechanism for global and local defect
formation at first order phase transitions.  We analyze the
equations of motion for scalar and gauge fields, and demonstrate the validity
of the geodesic rule for $\phi$.  To be specific, we consider vortex (in $2 +
1$ dimensions) or cosmic string (in $3 +1$ dimensions) formation in a model in
which a $U(1)$ global or local symmetry is broken.  However, our methods also
work for more complicated models with $\pi_1 ({\cal M}) \neq 1$, for other
types of topological defects, and for semilocal$^{\seventeen)}$ and
electroweak$^{18)}$ strings.  We do not here consider second order
transitions because the analysis is qualitatively different:  the semiclassical
methods we use ignore  thermal fluctuations.

In Section 2, we discuss theories with a first order phase transition which
procedes via bubble nucleation.  We discuss how far the analysis can be
applied to semilocal defects and electroweak strings in Section 3, and Section
4 contains a brief summary of results. Throughout the paper, units in which
$k_B = \hbar = c = 1$ are employed. Greek indices run from 0 to 3 and our
space-time signature is $(+, -, -, -)$.

\chapter{First order phase transitions}

To be specific, we shall consider an Abelian Higgs model with a complex scalar
field $\phi$ and a $U(1)$ gauge connection $A_\mu$.  The Lagrangian of the
system is
$$
{\cal L} = (D_\mu \phi)^{\dag} D^\mu \phi - V (\phi) - {1\over 4}  \,
F_{\mu\nu}
F^{\mu \nu} \, \eqno\eq
$$
where
$$
D_\mu = \partial_\mu - ie A_\mu \eqno\eq
$$
is the covariant derivative ($e$ is the gauge coupling constant),
$$
F_{\mu\nu} = \partial_\mu A_v - \partial_v A_\mu \eqno\eq
$$
is the field strength, and $V(\phi)$ is a symmetry breaking potential,
whose zero temperature form is $\la(|\phi|^2-\eta^2)^2/2$.  We
assume that $V(\phi)$ depends only on $|\phi|$.  In this case, the Lagrangian
(2.1) has a local $U(1)$ symmetry.  If $|\phi| = \eta(T) \neq 0$ is the
absolute
minimum of $V(\phi)$, then this symmetry is broken at low temperatures.  We
shall assume that $\phi = 0$ is the minimum of the high temperature effective
potential $V_T (\phi)$.  At a critical temperature $T_c$, $\phi = 0$ ceases
to be the absolute minimum of $V_T (\phi)$.

Considering the Abelian Higgs model in the context of an expanding Universe,
we conclude that as the temperature drops below $T_c$, the system undergoes a
symmetry breaking phase transition.  The order of the transition depends on
the functional form of $V_T (\varphi)$.  It is generally thought that the
transition is first order for $\beta \equiv \la/e^2 \to 0$, and may be
second order
for $\beta \to \infty.$$^{\MarRus)}$  In the former case, the one-loop
temperature dependent effective potential has the form$^{\EffPot)}$
$$
V_T(\phi) = {1\over 2} \la(T)(|\phi|^2-\eta^2)^2 + {1\over 4} e^2T^2
|\phi|^2 -  {\sqrt{2}\over 4\pi} e^3T|\phi|^3.
\eqno\eq
$$
In a first order phase transition, $\phi = 0$ remains a metastable fixed point
below $T = T_c$.  The transition to a state with $|\phi| = \eta$ procedes by
bubble nucleation.$^{\nineteen)}$  There is a finite probability per unit
volume per
unit time $dP/dVdt$ that a bubble with $|\phi| = \eta$ will nucleate in a
surrounding sea of ``false vacuum" $\phi = 0$.   This probability is given
by an expression of the form$^{\nineteen,\Lin)}$
$$
{dP\over dVdt} = A\exp-S_{\rm E}[\bar\phi,\bar A]
\eqno\eq
$$
where $\bar\phi$ and $\bar A$ are extrema of the Euclidean action $S_{\rm E}$.
This field configuration represents a tunnelling process.    At high
temperatures (high compared with the scalar and vector masses $\ms$ and $\mv$)
the action is effectively
three dimensional, and the tunnelling solution is spherically symmetric
about a point.$^{\Lin)}$
If the energy difference between the minima at $\phi=0$
and $|\phi|=\eta(T)$ is small compared with the height of the barrier between
them, then the solution is well approximated by a thin walled spherical
bubble of true vacuum inside the false one.$^{\nineteen)}$
The  width of the bubble
wall is approximately $\ms^{-1}$, and the radius of the bubble is of order
$V_{\rm b}/\ms\De V$, where $V_{\rm b}$ is the barrier height and $\De V$
is the difference is free energy between the two phases.
The bubble will then expand with a speed which depends on the interaction
between the wall and the rest of the hot matter in the universe.  We shall
assume that the couplings are small enough that we may take the limiting
velocity to be 1.
The expansion of the bubble is fueled by
the conversion of potential energy density $V(0)$ to wall kinetic and gradient
energy.
The phase transition is completed when neighboring bubbles collide and the
fraction of space with $|\phi| = \eta$ approaches unity.  The correlation
length $\xi$ for this transition is the mean separation of bubbles.  Implicit
in our analysis are some assumptions about time scales: we are taking the
expansion rate to be much longer than the nucleation rate per Hubble volume,
so that we are justified in taking the spacetime to be Minkowsi.  The
expansion of the universe then serves to reduce the temperture as a known
function of time.

According to the Kibble mechanism, there is a fixed probability $p$ of the
order 1 that a defect will form in any correlation volume $\xi^3$.  The exact
value of $p$ depends on the type of defect, i.e., on the topology of the
vacuum manifold ${\cal M}$ (see e.g., Ref. \twenty\ for a recent calculation
of $p$
for various models).

Let us illustrate the Kibble mechanism for our toy model, the Abelian Higgs
model.  The phase $\alpha$ of $\phi$ is assumed to take random
values in different
bubbles.  After two bubbles meet, then if we follow a line in space connecting
the centers of the two bubbles, $\alpha$ is assumed to interpolate
more or less smoothly between its
values in the two bubbles.  The second part of Kibble's argument states that
$\alpha (x)$ will follow a geodesic in ${\cal M}$ in order to minimize the
potential energy.

A vortex can form when three bubbles collide -- as illustrated in Fig. 1.  If
$\alpha_i, \, i = 1,2,3$ are the phases of $\phi$ in the three bubbles, then a
vortex will form if the sum of the phase differences $\alpha_2 - \alpha_1$ and
$\alpha_3 - \alpha_2$ exceeds $\pi$.  In this case, $\alpha (x)$ will run from
0 to $2 \pi$ as we go along the circle $\gamma$, i.e., the field configuration
has winding number 1.  There is then a topological obstruction to the scalar
field reaching the vacuum manifold everywhere in the region bounded by
the lines connecting the centres of the bubbles.  The field must vanish
somewhere, and this is where the vortex forms.

As discussed in Ref. \twentyone,
a bubble collision is a quite violent event.  In
order to justify the Kibble mechanism (both for global and local symmetry
breaking), we must follow the evolution of amplitude and phase of $\phi$ and
demonstrate that the geodesic rule is valid.  In order to do this, we will
write down the dynamical equations which follow from (2.1).  We first consider
a
global theory and discuss the evolution of amplitude and phase of $\phi$.
Then, we make the transition to a local theory and show that the gauge fields
do not strongly perturb the evolution of $\phi$.  This analysis is done in a
particular gauge.  However, the final result -- the winding number -- is gauge
invariant.

In Lorentz gauge, i.e., with the choice
$$
\partial_\mu A^\mu = 0 \, \eqno\eq
$$
the variational equations which follow from (2.1) are
$$
(\partial_\mu \partial^\mu - 2ieA_\mu\pa^\mu - e^2 A_\mu A^\mu) \, \phi + 2 \,
{\partial V\over{\partial |\phi|^2}} \phi = 0 \eqno\eq
$$
and
$$
\partial_\mu \partial^\mu A_v - 2 e^2 A_v |\phi|^2 = - ei \phi^\ast \dbw_v \phi
\, . \eqno\eq
$$
It is convenient to separate Eq. (2.7) into equations for the amplitude $\rho$
and phase $\alpha$ of $\phi$.  Inserting
$$
\phi = \rho e^{i \alpha}
\eqno\eq
$$
into (5), we obtain
$$
\partial^2 \rho - (\partial \alpha - eA)^2 \rho - e^2  A^2 \rho + 2 \,
{\partial V\over{\partial \rho^2}} \rho = 0 \eqno\eq $$ and
$$
\partial^2 \alpha + 2 (\partial^\mu \alpha - eA^\mu)\partial_\mu \rho \,
{1\over \rho} = 0 \, . \eqno\eq $$

The collision of two bubbles in the Abelian Higgs model was studied
numerically in Ref. \twentyone.  We now demonstrate that we can reproduce the
essential features of the collision process using the above equations.

We first consider a theory with global $U(1)$ symmetry and choose axes
such that the centers of the bubbles lie on the $x$ axis.   We can
additionally boost in the $(y,z)$ plane to a frame in which the bubbles
nucleate simultaneously$^{\twentyone)}$,
and translate in the $x$ direction so that they
collide at $x=0$.   Provided the bubbles nucleate far from each other
(far meaning much greater than the wall width so that the field is
exponentially close to zero between the bubbles) the solution representing
two bubbles nucleating at $\vec{x}_1$ and $\vec{x}_2$ can be approximated
by a sum ansatz
$$
\bar\phi(\vec x, 0) =  e^{i\al_1}f(\vec{x}-\vec{x}_1) + e^{i\al_2}f(\vec{x}
-\vec{x}_2)
\eqno\eq
$$
where $f$ is the modulus of the single bubble field. The action
for this configuration is (exponentially) independent of $\al_1$ and
$\al_2$:  hence the phases of the two fields are well approximated by
independent random variables.
Without loss of generality we take the phase in one
bubble to be $\alpha = 0$ and in the other bubble $\alpha = \alpha_c$.
Note that the phase $\alpha$ is not defined for points outside the bubbles.

When the bubbles meet (we take this to occur at time $t=0$), the phase
$\alpha(x)$ is approximately a step function
$$
\alpha (x, t = 0) = \alpha_c \theta (x) \, . \eqno\eq
$$
In fact, the step will be smoothed on a scale $\ms^{-1}$ by the finite
thickness of the bubble walls. We shall work in the planar approximation
$\partial_y \phi = \partial_z \phi =
0$. This is reasonable if the bubble radius is much greater than the wall
thickness.  In fact, the symmetry of the two bubble collision reduces the
problem to a two dimensional one anyway$^{\twentyone)}$,
but for simplicity we choose
not to exploit the coordinates in which this is manifest. Inserting (2.13) as
initial condition into the phase equation (Eq. (2.11)), we see that phase wave
fronts emerge which travel in $\pm x$ direction with the speed of light (see
Fig. 2): $$ \alpha (x, t \ge 0) = {1\over 2} \alpha_c \, [\theta (x + t) +
\theta (x-t)] \, . \eqno\eq $$
Eq. (2.14) gives the solution of (2.11) because the
phase waves are propagating inside the bubbles where $\partial_\mu \rho = 0$.

The phase waves which arise for $\alpha_c \ne 0$ carry away some of the kinetic
energy of the walls, but not all.  The rest of the energy goes into bubble
walls.  The region of false vacuum $\phi = 0$ does not disappear at $t = 0$.
Rather, the walls which separate the region with $|\phi| = \eta$ from the
false vacuum scatter and start to reexpand$^{\twentyone)}$.  What is happening
is that
the modulus of the field has overshot $\rho=\eta$, and rolled back to
$\rho=0$.  We denote the wall positions by $\pm X (t)$. However, as $X(t)$
increases, the potential energy of the field configuration increases, thus
creating a force which tends to restore $X$ to 0.  An approximate equation
for $X (t)$ can be obtained from elementary physics considerations: $$
\si \ddot X = - {\partial W\over{\partial X}} \, , \eqno\eq $$ where $\si$ is
the mass per unit area of the wall, and $W (X)$ is the potential energy (per
 unit area) of the
false vacuum region $x \in [-X, X]$.  Obviously, $ W = X
V(0)$ and hence (2.13) becomes
$$ \ddot X = - {V (0)/ \si}
 \eqno\eq $$
Thus, there is a constant restoring force which causes the wall to recollapse
on a time scale $  \tau = {2 \dot X (0)\si/ V(0)} \, .  $ There is a series of
bounces, each losing energy through propagating oscillations in $\rho$,
and thus leading to reduced $\dot X(0)$ and period $\tau$ (see Fig. 2).

An alternate way to derive Eq. (2.15) is to insert the ansatz
$$
\rho (x,t) = \eta \theta (x - X (t)) \eqno\eq $$
into the equation for the modulus $\rho$ (Eq. (2.10)), and integrate the
resulting distributional equation over $x$.

To summarize, we have verified that after a bubble collision, the
phase $\alpha (x)$ along a path $\gamma$ connecting the two bubble
centers interpolates between its original values in the two bubbles.
The interpolation happens in two jumps of width $\ms^{-1}$ associated with
excitations of the Nambu-Goldstone mode, spreading from the collision site
at the speed of light.  Thus, the
second key ingredient of the Kibble mechanism, the ``geodesic rule,"
has been established for global defects forming in a first order phase
transition.  We stress that the geodesic rule follows from the equations of
motion, not from minimizing the energy, as assumed in Ref. \sixteen.

We now extend the analysis to theories with a local $U(1)$ symmetry.
We again study the collision of two bubbles and establish the
applicability of the geodesic rule.  In the bubble nucleation instanton, the
magnetic field associated with $A_\mu$ vanishes.  Thus at the time of
nucleation $\tn$ we have, even in the Lorentz gauge, some freedom left in
specifying $A_\mu$.  In general,
$$
\vec{A}(\vec{x},\tn) = \na\La(\vec{x},\tn), \qquad A^0 = \dot\La(\vec{x},\tn),
\eqno\eq
$$
with $\La(\vec{x},\tn)$ an arbitrary function of position.  However, it is
clearly simplest to choose $\La(\vec{x},\tn)=0=\dot\La(\vec x,\tn)$,
which amounts to a complete
specification of the gauge.
During the bubble wall collision, a nonvanishing gauge connection is generated
through the coupling to the phase difference in the scalar field across the
bubble wall, which in turn feeds back into the evolution of the scalar field
modulus $\rho$ via (2.10). We will analyze firstly assuming $\la/e^2 \gg 1$.
We assume  that we may neglect $eA$ in comparison to $\pa\al \sim
\sqrt{\la}\eta$, so that Eq. (2.8) becomes
$$ (\pa_t^2   - \pa_x^2)A_{ v} = 2 e |\phi|^2 \, (\partial_v \alpha - e
A_v) \simeq 2 e \eta^2 \partial_v \alpha \, .  \eqno\eq $$
If we take for $\al$ Eq. (2.14) then we can easily solve for the gauge
field. This is most easily expressed in light cone coordinates $x^\pm = t\pm
x$.  After one integration we find ($t>0$)
$$ \eqalign{
\pa_-A_+ &= \frac{1}{4} e\eta^2\al_c \th(x^+), \cr
\pa_+A_- &= -\frac{1}{4} e\eta^2\al_c\th(x^-),\cr}
\eqno\eq
$$
which corresponds to an electric field in the $x$ direction, of
$$
F_{01} = 2(\pa_-A_++\pa_+A_-) = \half e\eta^2\al_c[\th(x+t)-\th(x-t)].
\eqno\eq
$$
This field exists only between the phase waves, which are carrying off equal
and opposite charges (per unit area) $\al_c/2$.  Behind the phase wave  the
Higgs vacuum screens the charges, and the gauge field is exponentially
damped beyond lightcone distances $\De x^\pm \sim (e\eta)^{-1}$. Thus we
conclude that the gauge field is bounded in modulus by $\sim \eta$.  Our
approximation is self-consistent, for $e\eta \ll \sqrt{\la}\eta$. At the
phase wave itself the gauge field vanishes, so it should be a good
approximation to ignore its effect on the propagation of the wave (this
feature was originally found by Hawking {\it et al\/} in their numerical
simulations$^{\twentyone)}$).

This is perhaps not the correct limit to use if we are assuming a first
order phase transition, where $\la/e^2$ is supposed to be small.  In this
limit we cannot approximate the phase wave by a step function.  Instead we take
($t>0$)
$$
\al(x) = {1\over 2} \al_c[1+\half W(\ms x^+)+\half W(\ms x^-)]
\eqno\eq
$$
where $W$ is the phase wave profile, interpolating between -1 and 1 as its
argument changes from $-\infty$ to $+\infty$.  Then we find that Eq.
(2.8) becomes
$$
(4\pa_+\pa_- + \mv^2)A_\pm = \pm\half\al_c e\eta^2 \ms W'(\ms x^\pm).
\eqno\eq
$$
In a gradient expansion in powers of $\ms/\mv$, the first term is
$A_\pm \simeq \pa_\pm\al/e = \pm \al_c \ms W'(\ms x^\pm)/4$.  Thus the current
vanishes to this order, and the equations of motion for $\al$ and $\rho$ are
affected only at higher order in $\ms/\mv$.

Physically, what is happening is that in the  limit $\ms  \gg \mv$, the
wall collision contains enough high frequency modes to create longitudinal
gauge bosons.  In the opposite limit, the gauge bosons are too massive to
appear from scalar bosons of freqency $\sim \ms$:  instead, the gauge
field tracks the phase to ensure that the current vanishes.

In either case, the key point is that the interpolation
of the phase between the bubbles is not affected by the presence of $A_v$.
Hence, by chosing a particular gauge, we establish the geodesic rule for local
theories.
In a three bubble collision, the geodesic rule can be applied between each
pair of bubbles provided the collision happens before the advancing wall
of the other bubble reaches the collision point.  We can in fact boost
along the normal to the plane containing the centres of the three bubbles
so that they nucleate simultaneously, so this is a constraint on the
positions of the centres: they must form an acute triangle.
The winding number around a closed curve connecting these points,
along which the scalar field vanishes nowhere, is a gauge
invariant quantity. This is given by
$$
n_\gamma = -{i\over 4\pi}\oint \, {\phi^\ast \dbw_\mu \phi\over{|\phi|^2}} \,
ds^\mu \, . \eqno\eq $$
We can compute it in our chosen gauge and be
confident that the result is gauge invariant.

These results imply that in first order phase transitions, there is a finite
probability that a field configuration of nontrivial winding (and hence a
vortex) emerges during the collision of three bubbles.  This is true for both
local and global theories.
 If $\alpha_1 , \, \alpha_2$ and $\alpha_3$ are the phases of $\phi$ in
the three bubbles (without loss of generality $\alpha_1 < \alpha_2 <
\alpha_3$) and if $\al_1+\pi < \al_3 < \al_2+\pi$, then by the geodesic rule,
after bubble collision, $\alpha (x)$ will smoothly go from 0 to $2 \pi$ along
$\gamma$, yielding a configuration with winding number 1.  As we have seen,
the phases in the three bubbles are random, since the nucleation probability
is independent of their values, and so we get winding number 1 (with
probability
1/4.$^{\twenty)})$

\chapter{Extensions} \par
Our analysis has been based on dynamical rather than on topological
considerations.  Hence, the arguments may carry over to the case of
defects such as semilocal strings$^{\seventeen)}$ or electroweak
strings$^{\eightteen)}$,
whose stability or otherwise is dynamical in origin.

Key to our analysis in previous chapters was to establish the validity
of the geodesic rule, i.e., of the statement that after completion of
the phase transition, the phases of the Higgs fields along a line in
space connecting the centers of two initial bubbles will interpolate
smoothly between their initial values in the bubbles, thus forming
section of geodesics on the vacuum manifold ${\cal M}$.  The question
of determining the probability of defect formation then reduces to the
statistics problem of finding the probability that a closed geodesic
will have nonvanishing winding number.

We now generalize this approach to semilocal and electroweak strings.
We first
verify the geodesic rule, thus reducing the problem to a probability
problem.  However, the probability calculation will in general be much
harder than for topological defects.

Semilocal strings$^{\seventeen)}$ arise in models with a large global symmetry
group $G$ of which only subgroup $\Gl$ is gauged.  If this gauge group and
its unbroken subgroup $H_{\rm l}$ obey the usual topological condition
$\pi_1(\Gl/H_{\rm l}) \ne 0$, then stable string solutions exist only if the
scalar mass is small enough relative to the vector
mass.$^{\twentyfour,\twentyfive)}$  The
simplest example is the model of Ref. \seventeen:
a U(2) symmetry acts on a complex
scalar doublet $\Phi$, but only the U(1) generated by the identity matrix is
gauged.  When $\Phi$ gains an expectation value the remaining symmetry is a
global U(1).

The vacuum manifold of the theory is $S^3$, defined by $|\Phi|^2 = \eta^2$,
and it is fibred by the action of the local U(1) into a bundle which is
locally $S^2\times S^1$ -- the Hopf bundle.$^{\twentyfour,\twentyfive)}$
In the string
solution, the scalar field wraps around one of these fibres outside the
string, and vanishes at the origin.  In fact, it is simply an embedding of
the Nielsen-Olesen$^{\twentythree)}$ vortex in the full theory.
It is only stable if
the scalar mass is less than the vector mass.

 The equations of
motion of this theory are essentially the same as those of Section 2.  Both
components $\phi_1$ and $\phi_2$ of $\Phi$ satisfy Eq. (2.5):
$$
\pa^2\phi_i -2ieA_\mu\pa^\mu\phi_i - e^2 A^2 \phi_i + 2 \, {\partial
V\over{\partial | \phi | ^2}} \, \phi_i = 0 \> \> i = 1,2 \, , \eqno\eq $$
and the gauge field satisfies the analog of Eq. (2.8):
$$ \partial^2 A_v - 2 e^2 A_v (\phi^2_1 + \phi^2_2) = - ie (\phi^\ast_1 \dbw_v
 \, \phi_1 + \phi^\ast_2 \dbw_v \, \phi_2 ) \, . \eqno\eq
$$ In particular, the gauge connection $A_\mu$ does not mix $\phi_1$ and
$\phi_2$.  The same arguments as in Section 2  imply that the geodesic rule
applies, in our Lorentz gauge with $A_\mu(\vec{x},t_{\rm n})=0$.

However, this is a geodesic rule in the full vacuum manifold $S^3$, in which
there is no topological obstruction to the scalar field reaching $|\Phi|=\eta$
everywhere in the triangular region formed by a three-bubble collision.  This
is related to the fact that the analogue of (2.24), or
$$
m_\ga = -{i\over 4\pi}\oint {\Phi^{\dag}\dbw_\mu\Phi\over  |\Phi|^2}dS^\mu,
\eqno\eq $$
is not necessarily an integer.  The probability
of forming a semilocal string is in fact a complicated dynamical question.
It seems likely however, that the `closer' the field in the three bubbles lies
to the same U(1) orbit, the more likely is the formation of a string.  We can
in fact give this closeness a precise geometrical meaning.

Let us choose coordinates $(\si,\psi,\chi)$ on $S^3$ such that in its vacuum
manifold
$$
\Phi = \eta { \cos\chi/2\, e^{i(\si-\psi)/2} \choose \sin\chi/2\,
e^{i(\si+\psi)/2}}.
\eqno\eq
$$
Then $\si$ parameterizes the $S^1$ fibres, and $(\chi,\psi)$ are polar
coordinates on the base space $S^2$.  This can be seen by projecting onto a
unit 3-vector
$$
\hat\vp^a = \Phi^{\dag}\si^a\Phi/|\Phi|^2 =
(\sin\chi\cos\psi,\sin\chi\sin\psi,\cos\chi).
\eqno\eq
$$
The semilocal string has $(\chi,\psi)$ constant around it,
with $\si$ changing by
$4\pi$.  Thus the measure of closeness of the field values in the three
bubbles $(\si_i,\chi_i,\psi_i)$ to the string configuration is the area of
the spherical triangle defined by $(\chi_i,\psi_i)$.  The smaller this area,
the closer the field is to a pure phase change around the curve joining the
centres of the bubbles (that is, the closer $m_\ga$ is to 1), and the more
likely a string is to form. Unfortunately, without dynamical simulations, we
cannot say how the string formation probability depends on the area $A$.  The
only piece of information we can extract geometrically is the probability
$P(A)$ that a random spherical triangle has area less than or equal to $A$,
which is$^{\AreaProb)}$
$$
P(A) = \[{A\over 2} + {1\over 2}\sin{A\over 2}\cos{A\over 2} -
\pi\sin^2{A\over 2} + {1\over 8}(\pi-A)(3\pi-A)\tan{A\over
2}\]/\pi\cos^2{A\over 2}. \eqno\eq
$$
For small $A$,
$$
P(A) \simeq {3A\over 4\pi}\(1+{\pi^2\over 4}\).
\eqno\eq
$$
For fields $\Phi$ with $d$ components we would expect the probability to go as
$V$, where $V$ is a $2d-1$ dimensional volume, although the coefficient is a
more difficult exercise in geometric probability.

For electroweak strings the geometrical considerations are identical: our
two component semilocal model is just the bosonic sector of the electroweeak
theory in the limit that the weak mixing angle approaches $\pi/2$.  However,
we can expect the fields to evolve towards a string solution only if there
is a locally stable solution towards which to evolve.  In the Standard Model
the string is unstable$^{\twentysix)}$, so we do not expect to see them formed.
However,  metastable  strings exist in theories with more complex Higgs
sector,$^{\HolPlus)}$
and in those we can expect some string formation in first order phase
transitions, although we are unable to estimate the probability.

\chapter{Conclusions}

We have studied the formation of topological defects in first
order phase transitions.  We showed that both for local and for
global defects, the assumptions on which the Kibble mechanism$^{\one)}$
is based
can be established by using the equations of motion.

The results of the analysis is that there is a probability $p$ that a
defect will form per correlation volume of the field.  For topological
defects, $p$ is of the order 1$^{\twenty)}$, for nontopological defects
such as semilocal strings and electroweak strings, $p$ depends on as yet
ill-understood dynamics, but we guess it to be typically much smaller than 1.
However, in no circumstance is the formation probability Boltzmann suppressed.

Our technique is based on first studying the dynamics of the scalar
fields in the absence of gauge fields (a gauge dependent analysis),
establishing the existence of winding number in the final state (a
gauge independent conclusion), and then bounding the effects of the
gauge fields, for convenience working in Lorentz gauge with the additional
choice   $A_\mu = 0$ at the time of bubble nucleation.

We have not included thermal fluctuations into our
consideration.  We are therefore implicitly assuming that the root mean
square amplitude of the thermal fluctuations on the length scales we discuss
is much less than the magnitude of the scalar field $\eta(T)$, which is
in fact not unreasonable.$^{\twentyseven)}$

\ack
For useful discussions and correspondence we are grateful to T.W.B.
Kibble and A. Srivastava.  We are also indebted to G.W. Gibbons for pointing
out reference \AreaProb.  This work was supported in part by DOE
contract DE-FG02-91ER40688, Task A, and by an NSF-SERC Collaborative
Research Award NSF INT-9022895, and by SERC GR/G37149.

\endpage

\baselineskip=19pt
\refout

\endpage

\chapter{Figure captions}

{\bf Figure 1}

Three bubbles of the broken symmetry phase ($\rho=\eta$) colliding.
If the phase change of the scalar field around the loop $\gamma$ is
$\pm 2\pi$, a string (or antistring) is formed.  If the phases
$\al_i$ are ordered, then then the requirement for a string is
$\al_1+\pi < \al_3 < \al_2+\pi$.

\bigskip
{\bf Figure 2}

Space-time diagram of two bubbles with different phases colliding.
After nucleation at
time $\tn$ the bubble walls collide at $t=0$, when they are travelling
at approximately $c$.  Phase waves continue out from the collision site,
spreading a region whose phase is halfway between the phases of
the bubbles.  The walls pass through each other, but eventually turn
round and recollide.  This may happen several times.

\bye